\documentclass[preprint,2.5p,times]{elsarticle}

\usepackage[utf8]{inputenc}
\usepackage{amsmath,amssymb,amsfonts}
\usepackage{graphicx}
\usepackage{braket}
\usepackage{xcolor}
\usepackage{hyperref}
\usepackage{subcaption}
\usepackage{lineno}
\usepackage[ruled,vlined,linesnumbered]{algorithm2e}
\usepackage{mathtools}
\biboptions{numbers,sort&compress}

\DeclareMathOperator{\Tr}{Tr}
\makeatletter
\def\ps@pprintTitle{%
  \let\@oddhead\@empty
  \let\@evenhead\@empty
  \def\@oddfoot{\centerline{\thepage}}%
  \let\@evenfoot\@oddfoot
}
\makeatother
\begin{document}

\begin{frontmatter}

\title{One-Sided Device-Independent Quantum Key Distribution over Noisy Metropolitan Links: Noise Thresholds and Purification-Assisted Recovery}

\author[qatar]{Syed M. Arslan\corref{cor1}}
\ead{syan54988@hbku.edu.qa}
\author[qatar]{Muhammad T. Rahim}
\author[qatar]{Asad Ali}
\author[qatar]{Hashir Kuniyil}
\author[qatar]{Saif Al-Kuwari}

\cortext[cor1]{Corresponding author}

\address[qatar]{Qatar Center for Quantum Computing, College of Science and Engineering, Hamad Bin Khalifa University, Doha, Qatar}

\begin{abstract}
One-sided device-independent quantum key distribution (1SDI-QKD) offers a practical middle ground between fully device-independent protocols and standard QKD, achieving security with detection efficiencies as low as 50.1\% on the untrusted side. However, prior analyses assumed idealized channels. This work extends the 1SDI-QKD framework to amplitude damping, dephasing, and depolarizing noise. Results reveal a clear noise hierarchy: dephasing tolerates the most noise before security is lost ($p_{\text{crit}} \approx 0.50$ versus 0.449 and 0.220 respectively), while depolarizing noise imposes the most stringent detection efficiency requirements, exceeding 97\% at 20\% noise where amplitude damping requires 76\%. Security is lost while substantial entanglement remains ($C \approx 0.67$--$0.74$), demonstrating that steering violation, not merely entanglement, determines 1SDI-QKD security. Integrating the BBPSSW purification protocol shows that it functions primarily as a range-extension mechanism: five rounds extend dephasing-limited operation from 10~km to 36~km at 90\% detection efficiency. Accounting for both the $2^n$ pair cost and the cumulative success probability, the effective rate peaks at $ 10^{-3}$ bits per initial pair for dephasing but only $ 10^{-6}$ for amplitude damping and depolarizing noise, identifying phase noise limited links as the regime where purification is worthwhile.
\end{abstract}

 \end{frontmatter}

\section{Introduction}
\label{sec:intro}

Quantum key distribution (QKD) allows two distant parties (Alice and Bob) to establish a shared secret key whose security is guaranteed by quantum physics rather than computational assumptions. Classical cryptographic protocols, which rely on computational hardness, are vulnerable to advances in algorithms and the advent of quantum computing \cite{shor1994algorithms,scarani2009security}. In contrast, QKD protocols, such as BB84  \cite{bennet1984quantum}, offer unconditional security under idealized device models. Practical implementations with attenuated laser sources,
for instance, occasionally emit multiphoton pulses that expose BB84 to
photon-number-splitting attacks, a loophole closed by the decoy-state
method \cite{lo2005decoy}. More generally, real-world devices, however, inevitably deviate from their ideal specifications, leading to potential security loopholes and side-channel attacks \cite{gerhardt2011full, makarov2006effects, zhao2008quantum, fung2008security}. In fact, practical QKD implementations have been compromised by attacks that exploit detector imperfections, as exemplified by bright-light hacking of commercial systems \cite{lydersen2010hacking}. These vulnerabilities motivate the development of QKD schemes that remain secure even when devices are uncharacterized and untrusted. 

Device-independent QKD (DI-QKD) \cite{ghoreishi2025future} addresses these device-trust issues by certifying security from the observed violation of a Bell inequality under minimal assumptions (quantum/no-signaling, independent random choices, and secure labs). In a typical DI-QKD protocol, devices are treated as black boxes, and a secret key can be established if the measurement correlations violate a Bell inequality by a sufficient margin
\cite{zhang2022device}. This approach is robust against a wide range of device flaws or malicious behavior. However, fully DI-QKD imposes significant requirements: detection efficiencies must be extremely high (often well above 80\%) and channel noise and losses minimal to avoid the detection loophole \cite{zapatero2023advances}. In practice, such conditions are difficult to achieve over long distances. Therefore, DI-QKD experiments were mostly limited to laboratory-scale distances (meters) despite significant advances \cite{zhang2022device}. 

An alternative approach to overcoming detector-side vulnerabilities is measurement-device-independent QKD (MDI-QKD), which removes the need for trust in detection by employing an untrusted intermediary to perform a Bell-state measurement on photons from Alice and Bob. MDI-QKD effectively closes all detector side channels and has been successfully demonstrated over long fiber distances, establishing it as a practical candidate for near-term QKD networks \cite{lo2012measurement}. Nonetheless, MDI-QKD still requires trusted source devices that emit the prescribed quantum states, and thus is not fully device independent in the Bell-test sense. In particular, the users must trust their state preparation, and no Bell inequality is violated during key exchange. Therefore, fully DI-QKD remains the ultimate goal for device-untrusted security, but its experimental feasibility is currently limited by stringent efficiency requirements. 

To bridge the gap between the strong security of DI-QKD and the practicality of standard QKD, one-sided device-independent (1SDI) QKD protocols have been developed. In 1SDI, only one of the two devices (e.g., Alice’s detection system) is assumed to be trusted and well-characterized, while the other device (Bob’s) is treated as a black box. This asymmetric trust significantly relaxes the security requirements: security can be certified through an EPR-steering test, in which Alice’s trusted measurements and the resulting correlations serve to verify the entanglement and ``steerability” of Bob’s outcomes. Branciard \emph{et al.} \cite{branciard2012one} proved that quantum steering (the phenomenon underlying 1SDI security) is both necessary and sufficient for secure key distribution in this setting.  Recent receiver-device-independent QKD protocols \cite{ioannou2022receiver} similarly assume asymmetric trust and have mainly been studied in prepare-and-measure settings. In this paper, we complement these by focusing on an entanglement-based scheme and explicitly quantifying noise effects and purification gains. 

By leveraging steering inequalities instead of full Bell inequalities, 1SDI-QKD tolerates much lower detector efficiency on Bob’s side. Recent theoretical result by Masini and Sarkar \cite{masini2026one} proposed a 1SDI protocol (with two measurement settings per party) that is secure against general coherent attacks with 50.1\% detection efficiency at Bob’s side, close to the fundamental limit for any two-setting DI protocol. Despite this substantial improvement, their security analyses of 1SDI-QKD have largely assumed idealized conditions aside from photon loss. In contrast, we incorporate realistic quantum noise models and show that secure key is lost well before entanglement vanishes, a security--entanglement gap that has not previously been quantified for 1SDI-QKD. These noise sources are prevalent in practical fiber-optic and free-space quantum links \cite{OptimizingQKD}, and their impact on two-qubit correlations,
including entanglement sudden death and revival under classical dephasing
channels, has been characterised in a range of settings~\cite{ali2023influence}. They
must be accounted for to accurately evaluate the viability of 1SDI-QKD in real
networks. Consequently, these quantum noises significantly raise the efficiency requirements, from $50.1\%$ in the noiseless case to between $65\%$ and
$76\%$ at $10\%$ noise, and above $97\%$ for depolarizing noise at $20\%$,
thereby providing a more realistic assessment of 1SDI-QKD performance in real-world scenarios. As a countermeasure, we integrate the BBPSSW entanglement purification protocol \cite{bennett1996purification} into the 1SDI framework, demonstrating that entanglement purification can effectively mitigate noise-induced efficiency demands on the untrusted party, which enables secure key rates in regimes that would otherwise yield no key generation.
 
\subsection{Contributions}

Our contributions in this paper can be summarized as follows:

\begin{enumerate}
\item \text{Noise characterization:} We quantify how amplitude damping, dephasing, and depolarizing channels affect secure key rates and detection efficiency thresholds, identifying parameter regimes enabling positive key generation.
\item \text{Security-entanglement gap:} We demonstrate that key rates vanish before entanglement does for amplitude damping ($\gamma \approx 0.449$ with $C \approx 0.74$) and depolarizing noise ($q \approx 0.220$ with $C \approx 0.67$), revealing that sufficient steering violation, not merely entanglement, determines security.
\item \text{Purification analysis:} We integrate the BBPSSW protocol and demonstrate its efficacy in restoring security, while providing resource overhead analysis quantifying its resource cost, showing that purification acts principally as a range-extension mechanism.
\item \textbf{Operational guidelines:} We provide practical recommendations for deploying 1SDI-QKD over noisy channels.
\end{enumerate}

\subsection{Organization}
The rest of the paper is organized as follows: in Section \ref{pre} establishes the
preliminaries: the asymptotic key-rate bound, the entropic quantities entering
it, and concurrence as our entanglement measure. Section \ref{sec:framework} describes the
1SDI-QKD protocol, trust assumptions, measurement settings, and security
certification via steering inequalities, and introduces the three quantum
noise channels studied. Section \ref{sec:results} reports simulation results for secure key rates under varying noise and detector efficiencies, as well as the effects of initial entanglement. Section \ref{purification} evaluates the efficacy of entanglement purification (BBPSSW) in improving fidelity and key rates. We then provide a Conclusion in section \ref{conc}, summarizing our contributions, a brief Outlook on experimental implementation and future work.

\begin{figure*}[t]
    \centering
    \includegraphics[width=\linewidth]{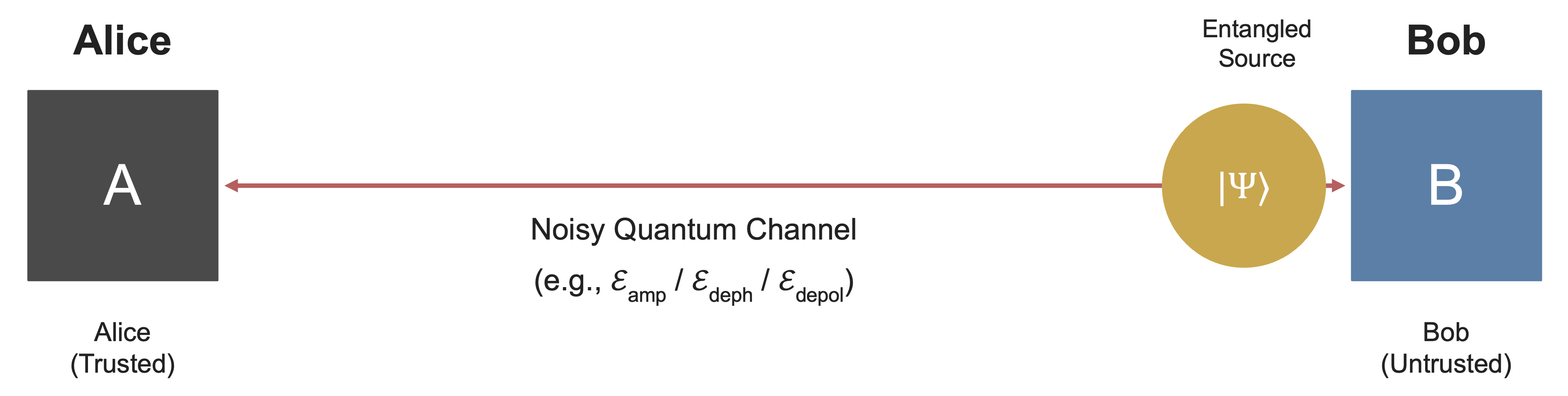}
    \caption{Schematic of the one-sided device-independent QKD setup. An entangled source prepares the two-qubit state $|\Psi\rangle$ and distributes one qubit to Alice (trusted) and the other to Bob (untrusted) through a noisy quantum channel characterized by amplitude-damping $\mathcal{E}_{\mathrm{amp}}$, dephasing $\mathcal{E}_{\mathrm{deph}}$, or depolarizing $\mathcal{E}_{\mathrm{depol}}$ noise.}
    \label{fig:1sdiqkd}
\end{figure*}

\section{Preliminaries} \label{pre}

In this section, we establish the mathematical framework for analyzing one-sided device-independent QKD security. We define the entropy measures that quantify information leakage, introduce concurrence as our entanglement metric, and establish the notation used throughout this paper.

\subsection{Asymptotic Key Rate}
The security of QKD protocols is quantified by the secret key rate, the number of secure key bits extractable per quantum signal. In the asymptotic limit of infinitely many signals, the Devetak-Winter bound~\cite{devetak2005distillation} establishes that the secure key rate satisfies
\begin{equation}
    r_\infty \geq H(A_1|E) - H(A_1|B_1),
    \label{eq:key_rate}
\end{equation}
where $H(A_1|E)$ quantifies Eve's uncertainty about Alice's measurement outcomes, and $H(A_1|B_1)$ represents Bob's uncertainty about Alice's outcomes. A positive value indicates that secure key bits can be distilled; when this quantity becomes negative, no secure key generation is possible. The two terms have distinct physical interpretations:
\begin{itemize}
    \item $H(A_1|E)$: conditional von Neumann entropy quantifying Eve's remaining uncertainty about Alice's raw key bits given her quantum side information. Higher values indicate greater secrecy and better security.
    \item $H(A_1|B_1)$: conditional Shannon entropy quantifying the classical correlation between Alice and Bob. Lower values indicate stronger correlations, benefiting key generation.
\end{itemize}

\subsection{Entropy}
The security of any QKD protocol ultimately rests on quantifying how much information about Alice's raw key is accessible to Bob versus to an eavesdropper Eve. In the asymptotic limit, the extractable secret key rate is given by the Devetak--Winter bound~\cite{devetak2005distillation},
\begin{equation}
    r = H(A_1|E) - H(A_1|B_1),
    \label{eq:dw}
\end{equation}
A positive key rate $r>0$ requires that Bob's knowledge of Alice's outcomes exceeds Eve's, that is, the legitimate parties share stronger correlations than any eavesdropper can obtain. We now discuss how each term is evaluated.

The conditional entropy between Alice's and Bob's measurement outcomes can be directly estimated from observed statistics. For measurement settings $x=1$ (Alice) and $y=1$ (Bob), it is defined as
\begin{equation}
    H(A_1|B_1) = -\sum_{a,b} p(a,b|1,1) \log_2 \frac{p(a,b|1,1)}{p(b|1)},
    \label{eq:hab}
\end{equation}
where $p(a,b|1,1)$ is the joint probability of Alice obtaining outcome $a$ and Bob obtaining outcome $b$ when both use measurement setting 1, and $p(b|1) = \sum_a p(a,b|1,1)$ is Bob's marginal probability. Equivalently, this can be written as
\begin{equation}
    H(A_1|B_1) = H(A_1,B_1) - H(B_1),
\end{equation}
where $H(A_1,B_1)$ is the joint entropy and $H(B_1)$ is Bob's marginal entropy.

Bounding Eve's uncertainty is more challenging since her attack strategy and quantum side information are unknown. In device-independent settings, we employ two complementary approaches:
\begin{enumerate}
\item {Analytical bounds.} For specific protocol structures, closed-form bounds can be derived relating $H(A_1|E)$ to observable quantities such as the steering parameter or CHSH value. Branciard \emph{et al.}~\cite{branciard2012one} established that for one-sided device-independent protocols, steering inequality violations directly bound Eve's information.

\item {Numerical optimization.} The Brown-Fawzi-Fawzi (BFF) method~\cite{brown2024device} provides tighter bounds by formulating $H(A_1|E)$ as a solution to a noncommutative polynomial optimization problem. This is implemented via the Navascu\'{e}s-Pironio-Ac\'{i}n (NPA) hierarchy~\cite{navascues2008convergent} of semidefinite programs, which yields a converging sequence of lower bounds on the conditional von Neumann entropy. In our analysis, we employ NPA level 1 with additional constraints reflecting Alice's measurement anti-commutativity.
\end{enumerate}

\subsection{Concurrence}
To track how channel noise degrades the shared quantum state, we employ Wootters' concurrence~\cite{wootters1998entanglement} as our entanglement measure. For a two-qubit density matrix $\rho$, define the spin-flipped state
\begin{equation}
    \tilde{\rho} = (\sigma_y \otimes \sigma_y) \rho^* (\sigma_y \otimes \sigma_y),
\end{equation}
where $\rho^*$ denotes complex conjugation in the computational basis. Let $\{\lambda_i\}_{i=1}^{4}$ be the eigenvalues of the product $\rho \tilde{\rho}$ arranged in non-increasing order. The concurrence is then
\begin{equation}
    C(\rho) = \max\left\{0, \sqrt{\lambda_1} - \sqrt{\lambda_2} - \sqrt{\lambda_3} - \sqrt{\lambda_4}\right\}.
    \label{eq:concurrence}
\end{equation}

Concurrence satisfies $0 \leq C(\rho) \leq 1$, vanishing for separable states and achieving unity for maximally entangled two-qubit states. As an entanglement monotone, it is non-increasing under local operations and classical communication (LOCC).

Concurrence provides a quantitative bridge between state quality and protocol security. However, a central finding of our work is that positive concurrence is \emph{necessary but not sufficient} for secure key generation: the state must violate the relevant steering inequality by a sufficient margin, which requires entanglement exceeding a threshold that depends on noise type. Tracking concurrence alongside key rates reveals where security fails despite residual entanglement, a phenomenon we call ``security-entanglement gap.''

Table~\ref{tab:notation} summarizes the key notation used throughout this paper.

\begin{table}[t]
\centering
\caption{Summary of notation.}
\label{tab:notation}
\begin{tabular}{cl}
\hline
\textbf{Symbol} & \textbf{Description} \\
\hline
$r_\infty$ & Asymptotic secure key rate (bits/pair) \\
$\eta_B$ & Bob's detection efficiency \\
$H(A_1|B_1)$ & Bob's uncertainty about Alice's outcomes \\
$H(A_1|E)$ & Eve's uncertainty about Alice's outcomes \\
$C(\rho)$ & Concurrence of state $\rho$ \\
$p$ & Dephasing probability \\
$\gamma$ & Amplitude damping parameter \\
$q$ & Depolarizing probability \\
$F$ & Fidelity with target Bell state $|\Phi^+\rangle$ \\
$L$ & Fiber length (km) \\
$L_c$ & Coherence length (km) \\
\hline
\end{tabular}
\end{table}

\section{One-Sided Device-Independent QKD} \label{sec:framework}

In this section, we describe the one-sided device-independent QKD protocol analyzed in this work. Our work is based on the BBM92 entanglement-based scheme~\cite{bennett1992quantum}, the entanglement-based analog of BB84, adapted to the one-sided device-independent setting following Branciard \emph{et al.}~\cite{branciard2012one} and incorporating recent efficiency improvements of Masini and Sarkar~\cite{masini2026one}, which we will refer to as \textit{1SDI-BBM92}. We detail the protocol assumptions, measurement procedures, security certification via steering inequalities, and the quantum noise models considered.

One-sided device independence is motivated by the asymmetric trust structures that arise naturally in practical deployment scenarios. Consider, for instance, a bank or government server that maintains well-characterized laboratory equipment with trusted detectors (Alice), communicating with a mobile client or remote user who employs commodity hardware that may be compromised, poorly calibrated, or supplied by an untrusted vendor (Bob). By removing trust assumptions on Bob's device, the protocol remains secure even if his detectors are manipulated by an adversary, a significant advantage over standard BBM92.


The original BBM92 protocol~\cite{bennett1992quantum}, introduced by Bennett, Brassard, and Mermin in 1992, showed that entanglement-based QKD could achieve equivalent security to prepare-and-measure BB84~\cite{bennett1984quantum} without requiring Bell inequality violation. Instead, BBM92 relies on the monogamy of entanglement: if Alice and Bob share a maximally entangled state, no third party (Eve) can be correlated with their measurement outcomes.

In standard BBM92, both parties possess trusted, well-characterized
measurement devices. The 1SDI-BBM92 protocol modifies this assumption
to achieve \emph{one-sided device independence}, as follows
(Table~\ref{protocol1} provides a detailed description):
\begin{itemize}
    \item \text{Alice (trusted):} possesses a fully characterized measurement device. Her detectors and measurement bases are assumed to operate as specified, as in standard BBM92.
    \item \text{Bob (untrusted):} his measurement device is treated as a black box. No assumptions are made about its internal workings; security must be certified from observed correlations alone.
    \item \text{Source:} an entangled photon source prepares two-qubit states and distributes one qubit to each party. We assume the source is located near Bob, so Alice's qubit traverses the noisy quantum channel.
\end{itemize}

This hybrid approach, combining the practical two-setting measurement structure of BBM92 with one-sided device independence, offers a middle ground between fully trusted QKD (vulnerable to detector attacks) and fully device-independent QKD (requiring impractically high detection efficiencies). Security certification shifts from device characterization to observable steering correlations.

\begin{table}[h]
\centering
\caption{Protocol 1: One-Sided Device-Independent QKD (1SDI-QKD)}
\label{protocol1}
\begin{tabular}{|p{\columnwidth}|}
\hline
\textbf{Protocol 1: One-Sided Device-Independent QKD (1SDI-QKD)} \\
\hline

\textbf{Setup:} Alice and Bob share an authenticated classical channel and are connected by a quantum channel. An untrusted source distributes $N$ copies of a bipartite state $\rho_{AB}$ to Alice (trusted) and Bob (untrusted). \\[0.5em]

\textbf{Quantum Phase:} \\
1. For each round $i \in \{1,\ldots,N\}$: \\
\hspace{1em}(a) The source distributes systems to Alice and Bob. \\
\hspace{1em}(b) Alice chooses $x_i \in \{1,2\}$ according to a known distribution $p(x)$ and obtains $a_i \in \{1,2\}$. \\
\hspace{1em}(c) Bob chooses $y_i \in \{1,2\}$ according to a known distribution $p(y)$ and obtains \\
\hspace{2em}an outcome $b_i \in \{1,2,\varnothing\}$ together with a detection flag $D_i \in \{\checkmark,\perp\}$. \\
2. Alice and Bob publicly announce the measurement settings $\{x_i\}$ and $\{y_i\}$, and Bob announces $\{D_i\}$. \\
3. If $D_i=\perp$, round $i$ is discarded. Otherwise ($D_i=\checkmark$), round $i$ is kept. \\[0.5em]

\textbf{Sifting and Parameter Estimation:} \\
4. For each kept round $i$, Alice samples a random bit $T_i \in \{k,t\}$: \\
\hspace{1em}$\bullet$ If $T_i=k$ (key round), Alice and Bob store $(a_i,b_i,x_i,y_i)$. \\
\hspace{1em}$\bullet$ If $T_i=t$ (test round), Alice and Bob publicly reveal $(a_i,b_i)$ to estimate $\vec{p}=\{p(a,b|x,y)\}$. \\
5. Using the estimated statistics $\vec{p}$, they compute the asymptotic key-rate bound $r_\infty(\vec{p})$ \\
\hspace{1em}and decide whether to abort (if $r_\infty\le 0$) or continue (if $r_\infty>0$). \\[0.5em]

\textbf{Key Distillation:} \\
6. From the stored key rounds (typically those with $(x_i,y_i)=(1,1)$ and $D_i=\checkmark$), Alice's outcomes form her raw key and Bob's outcomes form his raw key. \\
7. They perform error correction to reconcile their raw keys. \\
8. They perform privacy amplification to distill the final secret key. \\
\hline
\end{tabular}
\end{table}

\subsection{Measurement}

Following the BBM92 structure, both parties measure in two mutually unbiased bases corresponding to the Pauli $\sigma_z$ and $\sigma_x$ observables.

\subsubsection{Alice's Measurements (Trusted)}

Alice performs projective measurements in two bases:
\begin{align}
    \text{Setting } x=1: \quad \mathcal{M}_{A}^{(1)} &= \{|0\rangle\langle 0|, |1\rangle\langle 1|\} \quad (\sigma_z \text{ basis}), \\
    \text{Setting } x=2: \quad \mathcal{M}_{A}^{(2)} &= \{|+\rangle\langle +|, |-\rangle\langle -|\} \quad (\sigma_x \text{ basis}),
\end{align}
where $|\pm\rangle = \frac{1}{\sqrt{2}}(|0\rangle \pm |1\rangle)$. 

Since Alice's device is trusted, we apply the \emph{fair-sampling assumption}: rounds where her detector fails to click are discarded. This is justified by her device characterization, which ensures non-detection events are random and uncorrelated with the quantum state.

\subsubsection{Bob's Measurements (Untrusted)}

Bob's device is treated as a black box producing outcomes correlated with his input setting. Crucially, we \emph{cannot} assume fair sampling on Bob's side; doing so would open the detection loophole, allowing an adversary to fake quantum correlations by selectively triggering or suppressing detection events.

To rigorously account for detector inefficiency, we model Bob's effective measurement as a three-outcome POVM. For detection efficiency $\eta_B$ and nominal basis $p \in \{z, x\}$, let $\Pi_{0|p}$ and $\Pi_{1|p}$ be the ideal projectors satisfying $\Pi_{0|p} + \Pi_{1|p} = \mathbb{I}$. Bob's effective POVM elements are:
\begin{equation}
    M_{0|p}^{(\eta)} = \eta_B \Pi_{0|p}, \quad 
    M_{1|p}^{(\eta)} = \eta_B \Pi_{1|p}, \quad 
    M_{\varnothing|p}^{(\eta)} = (1-\eta_B)\mathbb{I}.
    \label{eq:bob_povm}
\end{equation}
Completeness is readily verified: $M_{0|p}^{(\eta)} + M_{1|p}^{(\eta)} + M_{\varnothing|p}^{(\eta)} = \mathbb{I}$.

The no-detection outcome $\varnothing$ is explicitly included in all statistical analyses. This conservative treatment is essential for device-independent security: by accounting for all possible outcomes, we prevent an adversary from exploiting detection inefficiency to compromise the protocol.

\begin{figure}
    \centering
    \includegraphics[width=.75\linewidth]{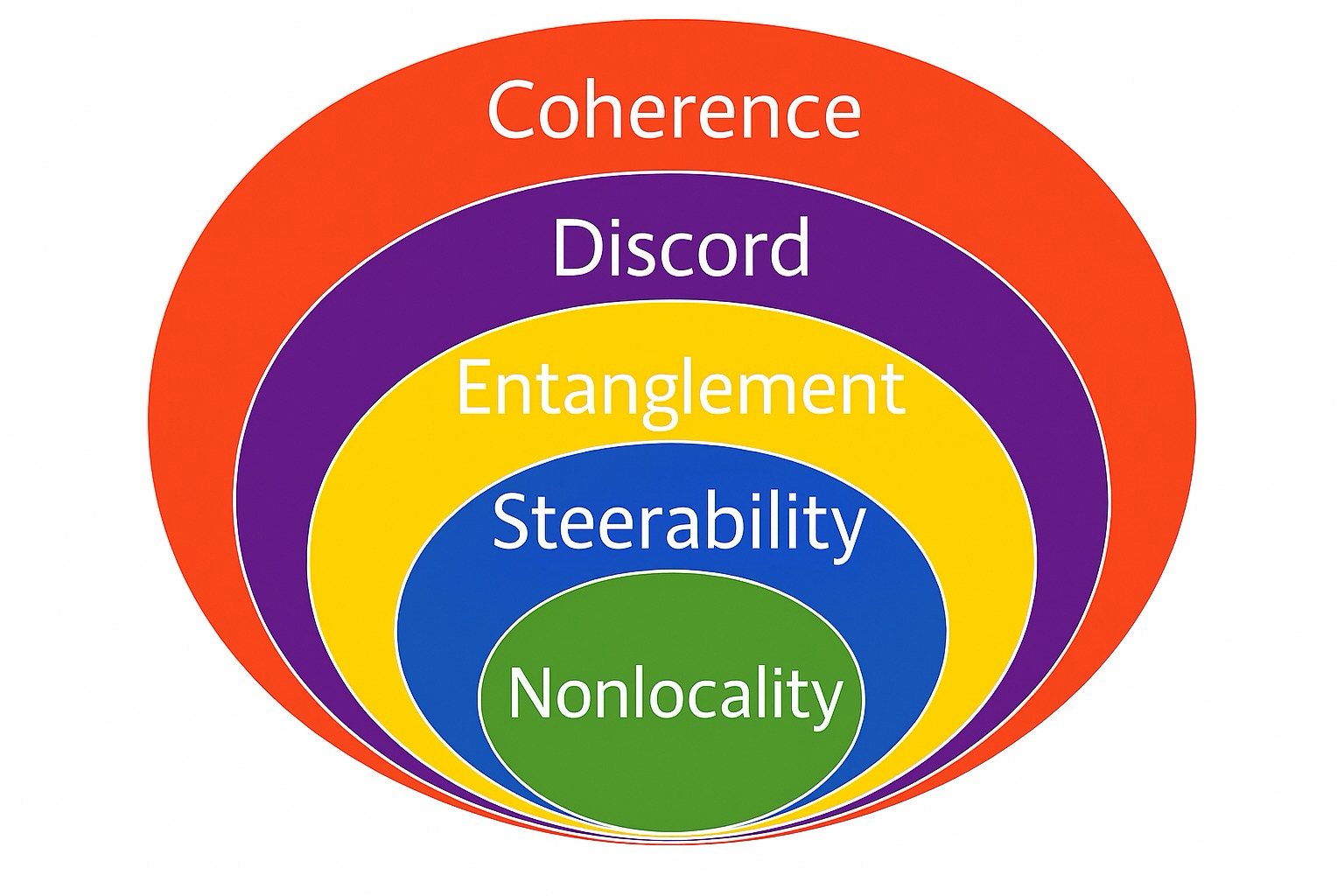}
    \caption{Hierarchy of quantum correlations. Nested sets illustrate the strict inclusion relations among different classes of quantum correlations \cite{wiseman2007steering}. 
    }
    \label{fig:hierarchy}
\end{figure}

\subsection{Security via Steering Inequality}

Unlike standard BBM92, which derives security from trusted device operation, the 1SDI-BBM92 certifies security through the violation of a \emph{steering inequality}, an observable test that cannot be passed by any local hidden state model. As shown in Fig.~\ref{fig:hierarchy}, quantum correlations form a strict hierarchy:
{\small
\begin{align*}
    \text{Nonlocality} \subset \text{Steerability} \subset \text{Entanglement} \subset \text{Discord}  \subset \text{Coherence}.
\end{align*}}

\noindent While fully device-independent QKD requires nonlocality (Bell violation), the 1SDI framework operates at the steerability level, a strictly intermediate resource between entanglement and nonlocality~\cite{wiseman2007steering,ali2024study}. States exhibiting nonlocality form the most restrictive class in the hierarchy of quantum correlations, while coherence represents the broadest quantum feature. By exploiting steerability, the 1SDI framework targets a weaker but more noise-robust requirement than Bell violation, making it resilient to higher channel noise than DI-QKD while providing stronger security guarantees than standard entanglement-based schemes that assume trusted devices.



Quantum steering~\cite{wiseman2007steering} is an asymmetric form of quantum nonlocality, intermediate between entanglement and Bell nonlocality. In a steering scenario, Alice (the trusted party) can verify that Bob's system is genuinely entangled with hers by demonstrating that the conditional states she can ``steer'' Bob's system into cannot be explained by any \emph{local hidden state} (LHS) model.
Branciard \emph{et al.}~\cite{branciard2012one} showed that steering is both \emph{necessary} and \emph{sufficient} for secure 1SDI-QKD:
\begin{itemize}
    \item \text{Necessary:} If the observed correlations admit an LHS model (no steering), then Eve could possess a copy of Bob's local hidden state, making secure key generation impossible.
    \item \text{Sufficient:} If steering is demonstrated, the correlations cannot be explained without genuine entanglement between Alice and Bob, guaranteeing that Eve's information is bounded.
\end{itemize}


For our BBM92-based protocol with two measurement settings per party, the relevant steering inequality is~\cite{cavalcanti2009experimental}:
\begin{equation}
    S_2 = \frac{1}{2}\left|\langle A_1 B_1 \rangle + \langle A_2 B_2 \rangle\right| \leq C_2^{\text{LHS}},
    \label{eq:steering_general}
\end{equation}
where $A_k$ and $B_k$ denote Alice's and Bob's measurement observables for setting $k$, and $C_2^{\text{LHS}}$ is the maximum value achievable by any LHS model.

With our specific choice of measurements ($A_1 = B_1 = \sigma_z$ and $A_2 = B_2 = \sigma_x$), the steering parameter becomes:
\begin{equation}
    S_2 = \frac{1}{2}\left(\langle \sigma_z \otimes \sigma_z \rangle + \langle \sigma_x \otimes \sigma_x \rangle\right),
    \label{eq:steering_2setting}
\end{equation}
with the LHS bound:
\begin{equation}
    S_2 \leq \frac{1}{\sqrt{2}} \approx 0.707.
    \label{eq:steering_bound}
\end{equation}

For the maximally entangled state $|\Phi^+\rangle$ with ideal measurements and detection, $\langle \sigma_z \otimes \sigma_z \rangle = \langle \sigma_x \otimes \sigma_x \rangle = 1$, yielding $S_2 = 1$, a maximal violation of the steering inequality. Noise and detection inefficiency reduce this value; the protocol remains secure only while $S_2 > 1/\sqrt{2}$.


The observed steering parameter directly constrains Eve's information about Alice's measurement outcomes. Following the analysis of Masini and Sarkar~\cite{masini2026one}, the conditional entropy $H(A_1|E)$ appearing in the key rate (Eq.~\ref{eq:key_rate}) is lower-bounded by a function of $S_2$ and the detection statistics:
\begin{equation}
    H(A_1|E) \geq f(S_2, \eta_B, \{p(a,b|x,y)\}),
\end{equation}
where the function $f$ is evaluated numerically using the BFF method~\cite{brown2024device} (see Section~\ref{sec:simulation}). This means that higher steering violations $\Rightarrow$ stronger quantum correlations $\Rightarrow$ larger $H(A_1|E)$ $\Rightarrow$ higher key rates. When $S_2$ falls below $1/\sqrt{2}$ due to noise, the correlations become classically explicable, $H(A_1|E)$ can no longer be guaranteed sufficient, and secure key generation becomes impossible.

\subsection{Noise Modeling}

In realistic fiber-optic or free-space implementations, the quantum channel introduces noise that degrades entanglement and reduces steering correlations. We model three prevalent noise types using the Kraus operator formalism \cite{nielsen2010quantum}. Unless stated otherwise, noise acts on Alice's qubit (which traverses the channel) via the map $\rho_{AB}' = (\mathcal{E} \otimes \mathcal{I})(\rho_{AB})$. We model one-sided channel noise on the traveling subsystem; for symmetric sources, this choice is equivalent up to relabeling the parties, and we apply $\mathcal{E}$ on Alice for notational convenience. We consider three noise models, as summarized in Table~\ref{tab:noise_summary}, and discussed below\footnote{In practice, multiple noise sources may act simultaneously. However, to isolate their individual impacts on 1SDI-QKD performance, we analyze each channel separately in Section~\ref{sec:results}. Combined noise scenarios can be treated by composing the respective channels.}. 

\subsubsection{Dephasing Channel}

Dephasing randomizes the relative phase between $|0\rangle$ and $|1\rangle$ without affecting state populations. It models:
\begin{itemize}
    \item Phase noise from fluctuating optical path lengths \cite{jones2018tuning}
    \item Magnetic field variations in atomic systems
    \item Timing jitter in photon detection
\end{itemize}

The Kraus operators are:
\begin{equation}
    K_0 = \sqrt{1-p}\,\mathbb{I}, \qquad K_1 = \sqrt{p}\,\sigma_z,
    \label{eq:dephasing_kraus}
\end{equation}
yielding the channel
\begin{equation}
    \mathcal{E}_{\text{deph}}(\rho) = (1-p)\rho + p\,\sigma_z \rho \sigma_z.
\end{equation}

In the computational basis, this damps off-diagonal elements by factor $(1-2p)$:
\begin{equation}
    \rho = \begin{pmatrix} \rho_{00} & \rho_{01} \\ \rho_{10} & \rho_{11} \end{pmatrix}
    \xrightarrow{\mathcal{E}_{\text{deph}}}
    \begin{pmatrix} \rho_{00} & (1-2p)\rho_{01} \\ (1-2p)\rho_{10} & \rho_{11} \end{pmatrix}.
\end{equation}

The dephasing probability relates to fiber length $L$ via \cite{kucera2024demonstration}:
\begin{equation}
    p(L) = \frac{1}{2}\left(1 - e^{-L/L_c}\right),
    \label{eq:dephasing_length}
\end{equation}
where $L_c$ is the coherence length characterizing the channel's phase stability. Note that $p \in [0, 0.5]$; at $p = 0.5$, complete dephasing occurs.

\subsubsection{Depolarizing Channel}

Depolarizing noise applies random Pauli errors with equal probability, uniformly shrinking the Bloch vector toward the maximally mixed state \cite{nielsen2010quantum}. It models:
\begin{itemize}
    \item Isotropic environmental coupling
    \item Polarization mode dispersion in fibers
    \item Multiple weak noise sources acting together
\end{itemize}

The Kraus operators are:
\begin{align}
    K_0 &= \sqrt{1-\frac{3q}{4}}\,\mathbb{I}, \quad
    K_1 = \sqrt{\frac{q}{4}}\,\sigma_x, \nonumber \\
    K_2 &= \sqrt{\frac{q}{4}}\,\sigma_y, \quad
    K_3 = \sqrt{\frac{q}{4}}\,\sigma_z,
    \label{eq:depol_kraus}
\end{align}
producing the channel
\begin{equation}
    \mathcal{E}_{\text{depol}}(\rho) = (1-q)\rho + q\frac{\mathbb{I}}{2}.
\end{equation}

This can be interpreted as: with probability $(1-q)$ the state is transmitted perfectly; with probability $q$ it is replaced by the maximally mixed state $\mathbb{I}/2$. The depolarizing probability varies with distance as:
\begin{equation}
    q(L) = 1 - e^{-L/L_c}.
    \label{eq:depol_length}
\end{equation}

\subsubsection{Amplitude Damping Channel}

Amplitude damping models energy relaxation processes where excited states decay to ground states. It describes:
\begin{itemize}
    \item Spontaneous emission in atomic systems
    \item Photon loss to the environment
    \item T$_1$ relaxation in superconducting qubits
\end{itemize}

The Kraus operators are:
\begin{equation}
    K_0 = \begin{pmatrix} 1 & 0 \\ 0 & \sqrt{1-\gamma} \end{pmatrix}, \qquad
    K_1 = \begin{pmatrix} 0 & \sqrt{\gamma} \\ 0 & 0 \end{pmatrix},
    \label{eq:amp_damp_kraus}
\end{equation}
yielding the transformation
\begin{equation}
    \mathcal{E}_{\text{ad}}(\rho) = \begin{pmatrix} 
        \rho_{00} + \gamma\rho_{11} & \sqrt{1-\gamma}\,\rho_{01} \\
        \sqrt{1-\gamma}\,\rho_{10} & (1-\gamma)\rho_{11}
    \end{pmatrix}.
    \label{eq:amp_damp_map}
\end{equation}

Unlike dephasing and depolarizing channels, amplitude damping is \emph{asymmetric}: it favors $|0\rangle$ over $|1\rangle$, breaking the symmetry of Bell-diagonal states. This has important consequences for entanglement purification, as discussed in Section~\ref{purification}.

The damping parameter scales with distance as:
\begin{equation}
    \gamma(L) = 1 - e^{-L/L_c}.
    \label{eq:amp_damp_length}
\end{equation}


\begin{table}[t]
\centering
\caption{Summary of quantum noise channels studied in this work.}
\label{tab:noise_summary}
\begin{tabular}{lccc}
\hline
\textbf{Channel} & \textbf{Parameter} & \textbf{Range} & \textbf{Physical Origin} \\
\hline
Dephasing & $p$ & $[0, 0.5]$ & Phase fluctuations \\
Depolarizing & $q$ & $[0, 1]$ & Isotropic noise \\
Amplitude damping & $\gamma$ & $[0, 1]$ & Energy relaxation \\
\hline
\end{tabular}
\end{table}

\section{1SDI-QKD Under Noise}
\label{sec:results}

In this section, we present simulation results characterizing 1SDI-BBM92 performance under realistic noise conditions.

\subsection{Simulation Methodology}
\label{sec:simulation}
To evaluate secure key rates under noise, we implement the following numerical procedure:
\begin{enumerate}
    \item \text{State preparation:} initialize the two-qubit state $\rho_0 = |\psi(\theta)\rangle\langle\psi(\theta)|$ with $\theta = \pi/4$ (maximally entangled) unless otherwise specified.
    \item \text{Noise application:} apply the relevant noise channel to Alice's qubit:
    \begin{equation}
        \rho_{AB} = (\mathcal{E} \otimes \mathcal{I})(\rho_0).
    \end{equation}
    \item \text{Measurement statistics:} compute joint probability distributions
    \begin{equation}
        p(a,b|x,y) = \text{Tr}\left[\left(M_a^{(x)} \otimes M_b^{(y,\eta_B)}\right) \rho_{AB}\right]
    \end{equation}
    for all measurement setting combinations $(x,y) \in \{1,2\}^2$ and outcomes $(a,b)$, incorporating Bob's detection efficiency via Eq.~\eqref{eq:bob_povm}.
    \item \text{Steering parameter:} calculate $S_2$ from the observed correlations using Eq.~\eqref{eq:steering_2setting}.
    
    \item \text{Entropy estimation:}
    \begin{itemize}
        \item $H(A_1|B_1)$: computed directly from $p(a,b|1,1)$ using Eq.~\eqref{eq:hab}.
        \item $H(A_1|E)$: lower-bounded using the Brown-Fawzi-Fawzi (BFF) method~\cite{brown2024device} with NPA hierarchy level 1, implemented via the Julia package
        with the MOSEK solver~\cite{mosek}.
    \end{itemize}
    
    \item \text{Key rate:} compute the asymptotic secure key rate:
    \begin{equation}
        r_\infty = H(A_1|E) - H(A_1|B_1).
    \end{equation}
    
    \item \text{Entanglement quantification:} calculate the concurrence $C(\rho_{AB})$ using Eq.~\eqref{eq:concurrence} to track entanglement degradation.
\end{enumerate}
All simulations use 100 efficiency values uniformly spaced in $\eta_B \in [0.4, 1.0]$ and noise
parameters sampled at intervals of 0.01 for threshold determination. For distance-dependent analyses in Section~\ref{purification}, we employ noise-type-specific coherence lengths that reflect the
distinct physical origins of each decoherence mechanism (see Section~\ref{purification} for details).

%
%
However, before analyzing noise effects, we validate our implementation by reproducing the fundamental efficiency threshold for 1SDI-QKD. In the absence of noise ($p = 0$), our simulations yield a minimum detection efficiency of $\eta_B^{\min} \approx 50.1\%$ for positive key generation, consistent with the theoretical prediction of Masini and Sarkar~\cite{masini2026one}. This threshold approaches the fundamental limit for any two-setting steering-based protocol, confirming the correctness of our BFF/NPA implementation.

All results presented in this section were obtained using:
\begin{itemize}
    \item BFF entropy bounds computed at NPA hierarchy level 1
    \item MOSEK solver at relative gap $10^{-6}$ and feasibility tolerance $10^{-7}$
    \item 100 uniformly spaced efficiency values in $\eta_B \in [0.4, 1.0]$
    \item Noise parameters sampled at intervals of 0.01 for threshold determination
\end{itemize}

The reported threshold values ($\gamma_{\text{crit}} \approx 0.449$, $q_{\text{crit}} \approx 0.220$) have numerical uncertainty of approximately $\pm 0.01$, limited by the noise parameter sampling resolution. 

In the following, we systematically analyze the impact of each noise type on secure key rates, detection efficiency, and entanglement degradation.

\begin{figure*}[t]
    \centering
    \begin{subfigure}[b]{0.45\textwidth}
        \centering
        \includegraphics[width=\textwidth]{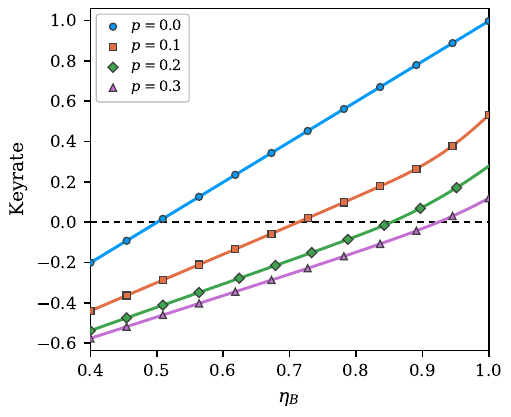}
        \caption{Dephasing}
        \label{fig:dephasing}
    \end{subfigure}
    \hfill
    \begin{subfigure}[b]{0.45\textwidth}
        \centering
        \includegraphics[width=\textwidth]{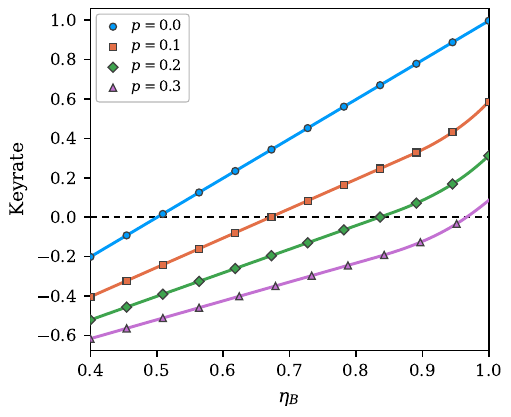}
        \caption{Amplitude damping}
        \label{fig:amplitude}
    \end{subfigure}
    \hfill
    \begin{subfigure}[b]{0.45\textwidth}
        \centering
        \includegraphics[width=\textwidth]{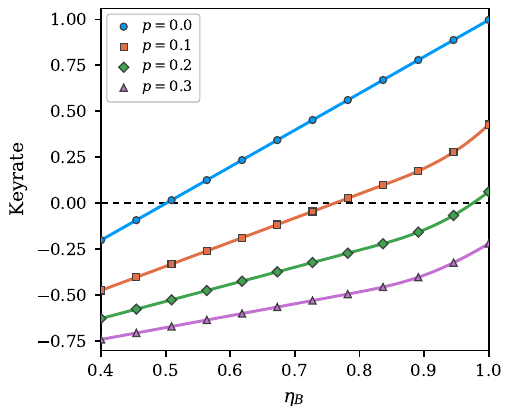}
        \caption{Depolarizing}
        \label{fig:depolarizing}
    \end{subfigure}
    \caption{Secure key rate $r$ versus Bob's detection efficiency $\eta_B$ under different noise models with for noise strengths 0.0 (blue), 0.1 (orange), 0.2 (green), and 0.3 (purple), where the noise strength is the dephasing probability $p$, the damping parameter $\gamma$, or the depolarizing probability $q$, respectively. (a)~Dephasing raises the minimum efficiency gradually, from $50.1\%$ without noise to $\approx 71\%$ at $p = 0.1$ and $\approx 85\%$ at $p = 0.2$, and positive key rates remain attainable at $p = 0.3$. (b)~Amplitude damping is the least demanding channel at low noise, requiring $\eta_B \approx 65\%$ at $\gamma = 0.1$ and $\approx 76\%$ at $\gamma = 0.2$, but the requirement rises steeply thereafter, exceeding $90\%$ at $\gamma = 0.3$. (c)~Depolarizing noise imposes the most stringent requirements, $\approx 76\%$ at $q = 0.1$ and $\approx 97.5\%$ at $q = 0.2$, with no positive key rate at any efficiency for $q = 0.3$.}
    \label{fig:keyrate_comparison}
\end{figure*}

\subsection{Secure Key Rates}
We investigate how each noise type affects the secure key rate $r$ of the 1SDI-QKD protocol. Unless stated otherwise, the initial shared state is maximally entangled ($\theta = \pi/4$), providing the highest baseline key rate.

\subsubsection{Detection Efficiency}

Figures \ref{fig:keyrate_comparison}
 show the secure key rate as a function of Bob's detection efficiency $\eta_B$ for varying noise levels. The three noise types exhibit markedly different behaviors:

\begin{enumerate}

\item {Dephasing noise:} dephasing raises the efficiency requirement gradually,
from $\eta_B^{\min} = 50.1\%$ in the noiseless case to $71\%$ at $p = 0.1$ and
$85\%$ at $p = 0.2$, remaining satisfiable throughout $p < 0.5$. This tolerance
arises because dephasing preserves the diagonal (population) elements of the
density matrix while degrading only coherences: the $\sigma_z$ measurement used
for key generation is insensitive to phase, so dephasing affects security
primarily through the reduced $\sigma_x$ correlations needed for steering
verification.

\item {Amplitude damping:} amplitude damping is the least demanding of
the three channels in the low-noise regime, requiring $\eta_B^{\min} \approx
65\%$ at $\gamma = 0.1$ and $76\%$ at $\gamma = 0.2$. Because the channel acts
on Alice's trusted qubit, it biases her marginal toward $|0\rangle$ and thereby
reduces $H(A_1|B_1)$, partly offsetting the loss of coherence through the
$\sqrt{1-\gamma}$ damping factor. The requirement nonetheless rises steeply
beyond $\gamma \approx 0.3$, and no efficiency suffices past
$\gamma_{\rm crit} \approx 0.449$.

\item {Depolarizing noise:} depolarizing noise imposes by far the most
stringent requirements, reaching $76\%$ at $q = 0.1$ and $97.5\%$ at $q = 0.2$,
with no positive key rate available at any efficiency for $q > 0.220$.
Depolarizing noise uniformly contracts the Bloch vector, degrading correlations
in all directions equally, so the error rate grows in both measurement bases and
leaves minimal margin for additional detection loss.
\end{enumerate}


\begin{figure}[t]
    \centering
     \includegraphics[width=0.70\columnwidth]{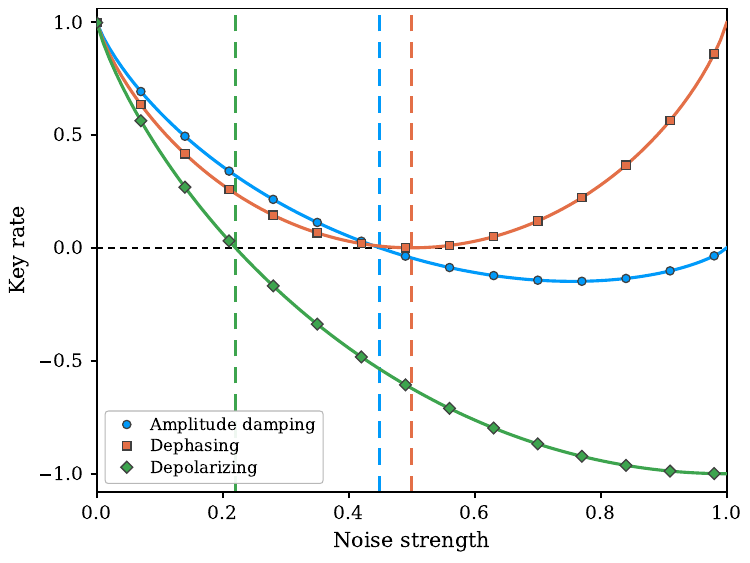}
    \caption{Secure key rate $r$ as a function of noise strength for amplitude damping (blue circles), dephasing (orange squares), and depolarizing (green diamonds) channels at ideal detection efficiency $\eta_B = 1$.}
    \label{fig:keyrate_vs_noise}
\end{figure}

To isolate the effect of noise from detection inefficiency, Fig.~\ref{fig:keyrate_vs_noise} plots the secure key rate as a function of noise parameter for each channel at ideal detection efficiency ($\eta_B = 1$). This reveals the intrinsic noise tolerance of the protocol: each noise type exhibits a critical threshold beyond which no secure key can be extracted, with $\gamma_{\text{crit}} \approx 0.449$ for amplitude damping, $q_{\text{crit}} \approx 0.220$ for depolarizing, and $p_{\text{crit}} \approx 0.50$ for dephasing (see also Table~\ref{tab:noise_thresholds}). The ordering $q_{\text{crit}} < \gamma_{\text{crit}} < p_{\text{crit}}$ establishes a clear \emph{noise hierarchy}: depolarizing noise is most detrimental, followed by amplitude damping, with dephasing being most tolerable.


\begin{table}[t]
\centering
\caption{Critical noise thresholds for 1SDI-QKD security. The key rate threshold $p_{\text{crit}}^{(r)}$ indicates where $r \to 0$ at $\eta_B = 1$. The steering threshold $p_{\text{crit}}^{(S)}$ indicates where $S_2$ reaches the classical bound. The gap between these thresholds reflects the additional penalty from increased $H(A_1|B_1)$.}
\label{tab:noise_thresholds}
\begin{tabular}{lcccc}
\hline
\textbf{Noise Type} & $p_{\text{crit}}^{(r)}$ & $p_{\text{crit}}^{(S)}$ & $\eta_B^{\min}$ at $p=0.1$ & Tolerance \\
\hline
Dephasing & $\approx 0.50$ & 0.293 & $\sim 71\%$ & High \\
Amplitude damping & $\approx 0.449$ & 0.375 & $\sim 65\%$ & Low \\
Depolarizing & $\approx 0.220$ & 0.293 & $\sim 76\%$ & Very low \\
\hline
\end{tabular}
\end{table}

These thresholds can be understood through the steering parameter $S_2$. For a maximally entangled initial state subjected to one-sided noise, the steering parameter evaluates to:
\begin{align}
S_2^{\text{deph}}(p) &= \frac{1}{2}(1 + |1-2p|), \\
S_2^{\text{depol}}(q) &= 1 - q, \\
S_2^{\text{ad}}(\gamma) &= \frac{1}{2}\left(1 - \gamma + \sqrt{1-\gamma}\right).
\end{align}

Security requires $S_2 > 1/\sqrt{2} \approx 0.707$. Solving for the critical noise values:
\begin{align}
p_{\text{crit}}^{\text{steering}} &= \left(1 - \frac{1}{\sqrt{2}}\right) \approx 0.293, \\
q_{\text{crit}}^{\text{steering}} &= 1 - \frac{1}{\sqrt{2}} \approx 0.293, \\
\gamma_{\text{crit}}^{\text{steering}} &\approx 0.375.
\end{align}

The numerically observed depolarizing thresholds ($q_{\text{crit}} \approx 0.220$) are \emph{lower} than corresponding steering-based estimate ($0.293$) because the key
rate depends on both $H(A_1|E)$ and $H(A_1|B_1)$: depolarizing noise degrades
the correlations in both measurement bases, so Bob's uncertainty $H(A_1|B_1)$
grows and the reconciliation cost exhausts the key rate before the steering
inequality is violated. For amplitude damping the ordering reverses
($\gamma_{\rm crit}\approx0.449$ vs.\ $0.375$): damping biases Alice's marginal
toward $|0\rangle$ and thus reduces rather than increases $H(A_1|B_1)$, so the
tighter numerical certification retains a positive rate beyond the point at
which the linear witness has ceased to respond. For dephasing the ordering reverses
($p_{\rm crit}\approx0.50$ vs.\ $0.293$): the $\sigma_z$ correlations used for
key generation are untouched, $H(A_1|B_1)$ imposes no penalty, and the tighter
certification retains a positive rate throughout $p<1/2$, where the linear
witness has already ceased to respond.

Table~\ref{tab:noise_thresholds} summarizes the critical thresholds and their practical implications. The clear noise hierarchy suggests prioritized mitigation strategies:
\begin{enumerate}
    \item \text{Depolarizing noise} (most harmful): Requires active polarization tracking and stabilization. Consider time-bin or phase encoding to avoid polarization-dependent errors.
    \item \text{Amplitude damping} (severe): Minimize transmission loss through high-quality optics. Position the source so that the trusted party's qubit traverses the shorter path, since channel loss on Alice's arm compounds with Bob's detection inefficiency.
    \item \text{Dephasing} (most tolerable): Standard interferometric stabilization suffices. Focus engineering efforts on the more damaging noise sources.
\end{enumerate}

\subsubsection{Entanglement Impact}

\begin{figure}[t]
    \centering
     \includegraphics[width=0.7\columnwidth]{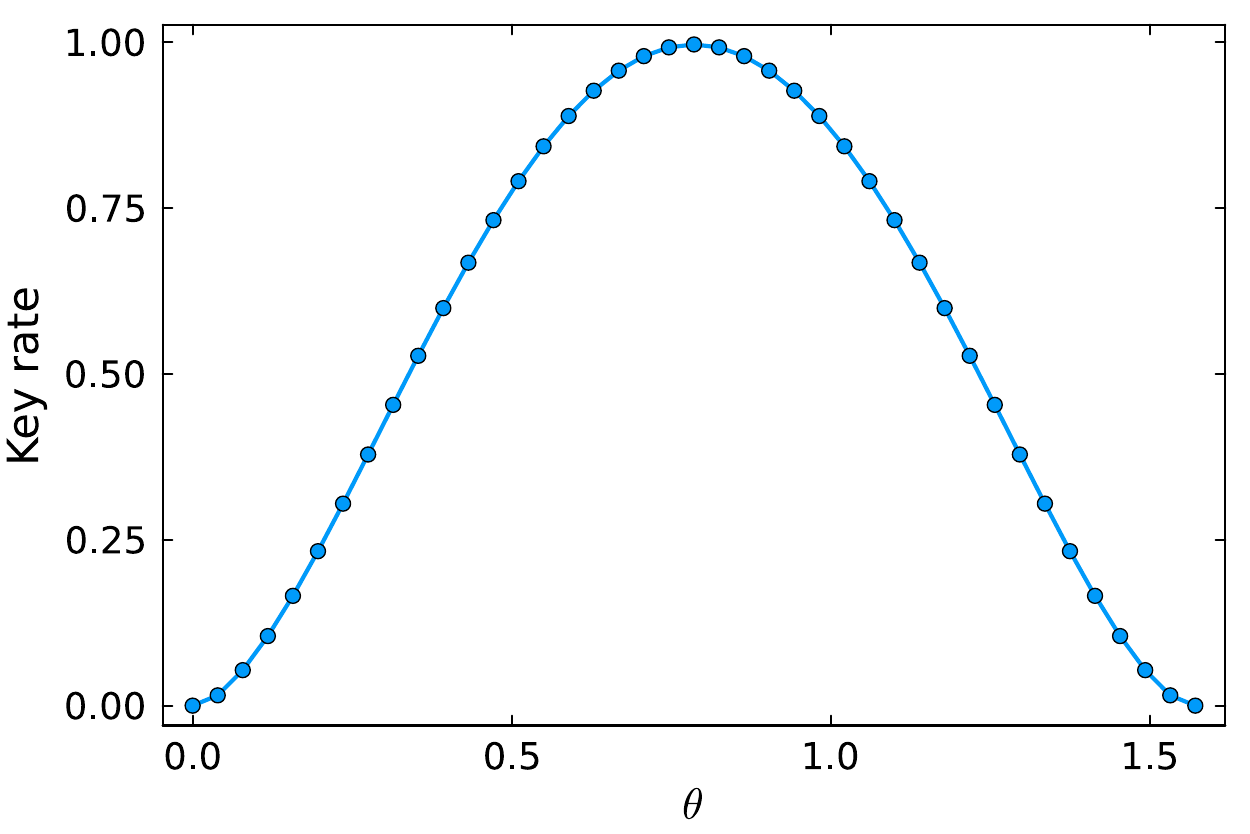}
    \caption{Secure key rate $r$ versus entanglement angle $\theta$ of the initial state $|\psi(\theta)\rangle$, assuming ideal detection efficiency ($\eta_B = 1$) and no channel noise. The maximum occurs at $\theta = \pi/4$ (maximally entangled state), with key rates vanishing as the state becomes separable.}
    \label{fig:keyrate_vs_theta}
\end{figure}

We next examine how the initial entanglement level affects protocol performance. Figure~\ref{fig:keyrate_vs_theta} shows the secure key rate versus the entanglement parameter $\theta$ for an ideal detector and no channel noise.

The maximum key rate occurs at $\theta = \pi/4$ (the maximally entangled Bell state $|\Phi^+\rangle$), as expected from theoretical considerations. The key rate drops symmetrically as $\theta$ deviates from $\pi/4$ in either direction, vanishing at $\theta = 0$ and $\theta = \pi/2$ (separable states). This dependence can be understood analytically. For the state $|\psi(\theta)\rangle = \cos\theta|00\rangle + \sin\theta|11\rangle$, the steering parameter evaluates to:
\begin{equation}
S_2(\theta) = \frac{1}{2}(1 + \sin 2\theta),
\end{equation}
which achieves its maximum of 1 at $\theta = \pi/4$ and falls below the classical bound $1/\sqrt{2}$ for $\theta < \theta_{\min}$ or $\theta > \pi/2 - \theta_{\min}$, where:
\begin{equation}
\theta_{\min} = \frac{1}{2}\arcsin\left(\sqrt{2} - 1\right) \approx 0.212~\text{rad} \approx 12.2^\circ.
\end{equation}


Under noise, highly entangled initial states generally outperform weakly entangled ones, providing a larger ``buffer'' against noise-induced degradation. A weakly entangled source ($\theta$ near 0.2 rad, concurrence $C \approx 0.38$) may lose all entanglement under moderate noise, while a maximally entangled source retains sufficient correlations for key generation. However, we observed a subtle effect with amplitude damping: as this channel preferentially decays the $|11\rangle$ component, states with $\theta > \pi/4$ (biased toward $|11\rangle$) suffer greater degradation than those with $\theta < \pi/4$. This asymmetry suggests that in amplitude-damping-dominated channels, slightly $|00\rangle$-biased states ($\theta$ slightly below $\pi/4$) might offer marginal advantages, though the effect is small in our simulations.

\subsection{Entanglement Degradation}
A central finding of this work is that secure key generation fails \emph{before} entanglement vanishes, a phenomenon we call the ``security-entanglement gap''. Understanding this gap is crucial for assessing when purification or other countermeasures are necessary.

Figure~\ref{fig:concurrence_vs_noise} shows the concurrence $C$ as a function of noise parameter for each channel, with noise applied to one qubit of an initially maximally entangled state. The three channels exhibit qualitatively different entanglement dynamics:

\begin{figure}[t]
    \centering
     \includegraphics[width=0.7\columnwidth]{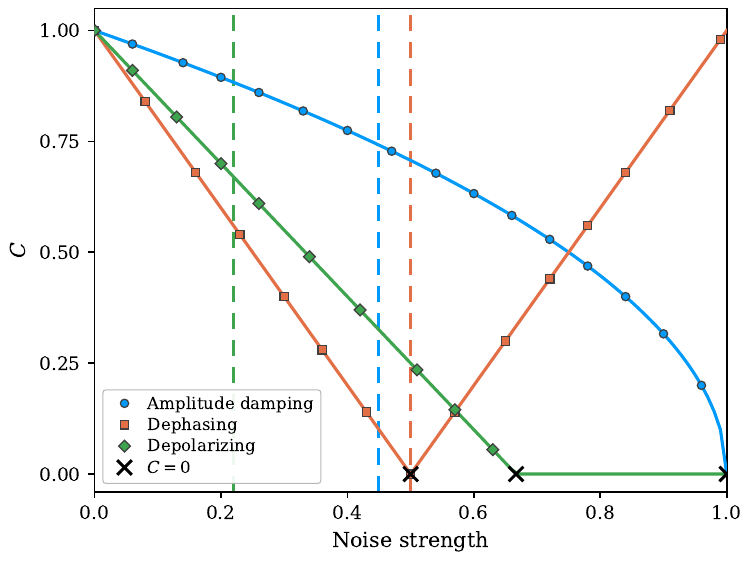}
    \caption{Concurrence $C$ versus noise strength for amplitude damping (blue,
$\gamma$), dephasing (orange, $p$), and depolarizing (green, $q$) applied to one
qubit of an initially maximally entangled state. Crosses mark zero-concurrence
points; only depolarizing noise shows entanglement sudden death, vanishing for
all $q\geq2/3$. Dashed vertical lines give the key-rate thresholds of
Table~\ref{tab:noise_thresholds}. Their separation from the corresponding
zero-crossings is the security--entanglement gap: security is lost while
$C\approx0.74$ (amplitude damping) and $C\approx0.67$ (depolarizing) remain.}
    \label{fig:concurrence_vs_noise}
\end{figure}

\begin{enumerate}

\item {Amplitude damping} Concurrence decreases continuously as:
\begin{equation}
C(\gamma) = \sqrt{1-\gamma},
\end{equation}
approaching zero only at complete damping ($\gamma = 1$). Notably, \emph{no entanglement sudden death occurs}, the decay is smooth and monotonic.

\item {Depolarizing noise:} Concurrence decreases linearly until reaching zero at the entanglement sudden death (ESD) point:
\begin{equation}
C(q) = \max\left\{0, 1 - \frac{3q}{2}\right\},
\end{equation}
yielding $q_{\text{ESD}} = 2/3 \approx 0.67$. Beyond this point, the state is separable.

\item {Dephasing noise} Concurrence exhibits non-monotonic behavior:
\begin{equation}
C(p) = |1 - 2p|,
\end{equation}
reaching zero at $p_{\text{ESD}} = 0.5$, then \emph{reviving} for $p > 0.5$ as the state transforms toward $|\Phi^-\rangle$. At $p = 1$ (complete phase flip), $C = 1$ again. The revival occurs because $p$ parametrises the \emph{probability} of a phase
flip rather than a decoherence strength: at $p=1$ the channel applies $\sigma_z$
deterministically, and being a local unitary it cannot reduce entanglement.
Decoherence is maximal at $p=0.5$, where the flip is least predictable, which is
why $C=|1-2p|$ depends only on the bias of the phase-flip probability. For practical purposes the relevant regime is $p<0.5$, since Eq.~\eqref{eq:dephasing_length}
confines the dephasing probability to $p\in[0,0.5]$ for any fiber length.
\end{enumerate}

\begin{table}[t]
\centering
\caption{Comparison of security and entanglement thresholds, demonstrating the security--entanglement gap. The residual concurrence $C$ at the key-rate threshold quantifies remaining entanglement when security is lost.}
\label{tab:security_entanglement_gap}
\begin{tabular}{lccc}
\hline
\textbf{Noise Type} & $\mathbf{p_{\mathrm{crit}}}$ ($r\!\to\!0$) & $\mathbf{p_{\mathrm{ESD}}}$ ($C\!\to\!0$) & $\mathbf{C}$ at $\mathbf{p_{\mathrm{crit}}}$ \\
\hline
Dephasing          & $\approx 0.50$ & $0.50$ & $\sim 0$ \\
Amplitude damping  & $\approx 0.449$ & $1.00$ & $\approx 0.74$ \\
Depolarizing       & $\approx 0.220$ & $0.67$ & $\approx 0.67$ \\
\hline
\end{tabular}
\end{table}


Table~\ref{tab:security_entanglement_gap} compares the key rate thresholds with the entanglement thresholds, revealing significant gaps: 
for amplitude damping and depolarizing noise, the key rate vanishes while substantial entanglement remains ($C \approx 0.67$--$0.74$). This demonstrates that \emph{entanglement is necessary but not sufficient for 1SDI-QKD security}. The protocol requires not merely nonzero entanglement, but sufficient entanglement to violate the steering inequality ($S_2 > 1/\sqrt{2}$), and maintain $H(A_1|E) > H(A_1|B_1)$ despite noise-induced correlation degradation. The only exception is dephasing, where security and entanglement thresholds coincide at $p = 0.5$. This occurs because dephasing affects only off-diagonal elements, and both steering violation and entanglement depend on the same coherence terms.


The security-entanglement gap has important practical implications, including:
\begin{enumerate}
    \item \text{Entanglement witnesses are insufficient:} simply verifying that a shared state is entangled does not guarantee secure key generation. Steering-based certification is essential.
    \item \text{Purification timing:} entanglement purification should be initiated when key rates approach zero, not when entanglement approaches zero. Waiting until ESD wastes resources on states that cannot yield secure keys.
    \item \text{Resource allocation:} the residual entanglement at $p_{\text{crit}}$ ($C \approx 0.67$--$0.74$) indicates that purification protocols, which typically require $F > 0.5$ (equivalently $C > 0$), remain viable even at the security threshold.
\end{enumerate}



\begin{figure}[t]
    \centering
    \includegraphics[width=.52\columnwidth]{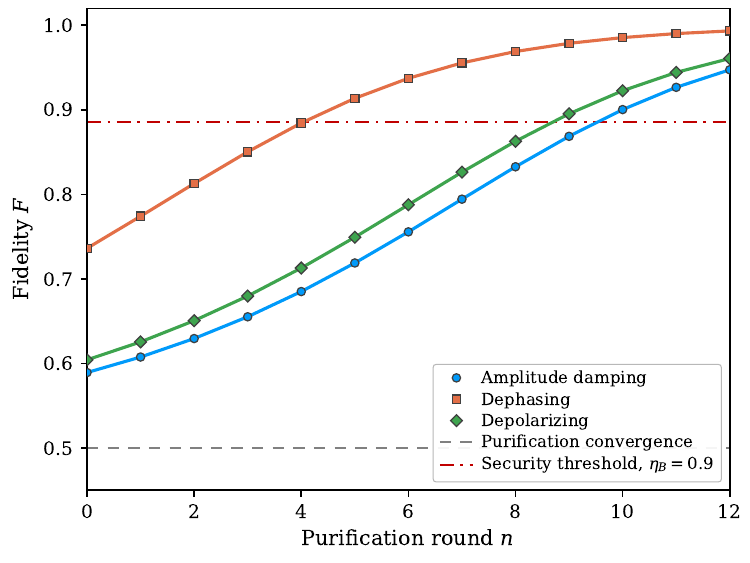}
    \caption{Fidelity versus purification round for dephasing ($L_c = 40$~km), amplitude damping ($L_c = 24$~km), and depolarizing ($L_c = 40$~km) noise at $L = 30$~km. The dashed line indicates the purification threshold $F = 0.5$ and security threshold at F = 0.885}
    \label{fig:fidelity_purification}
\end{figure}

\begin{figure*}[t]
    \centering
    \begin{subfigure}{0.48\textwidth}
        \centering
        \includegraphics[width=\linewidth]{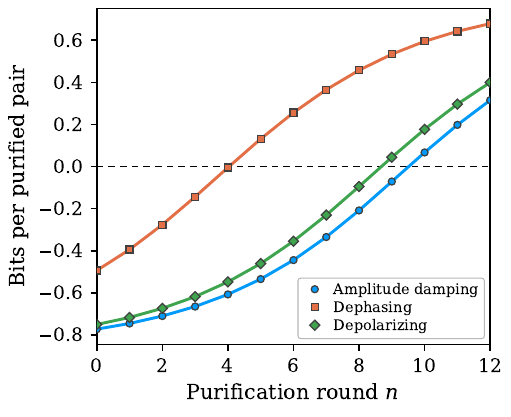}
        \caption{}
        \label{fig:keyrate_purification}
    \end{subfigure}
    \hfill
    \begin{subfigure}{0.48\textwidth}
        \centering
        \includegraphics[width=\linewidth]{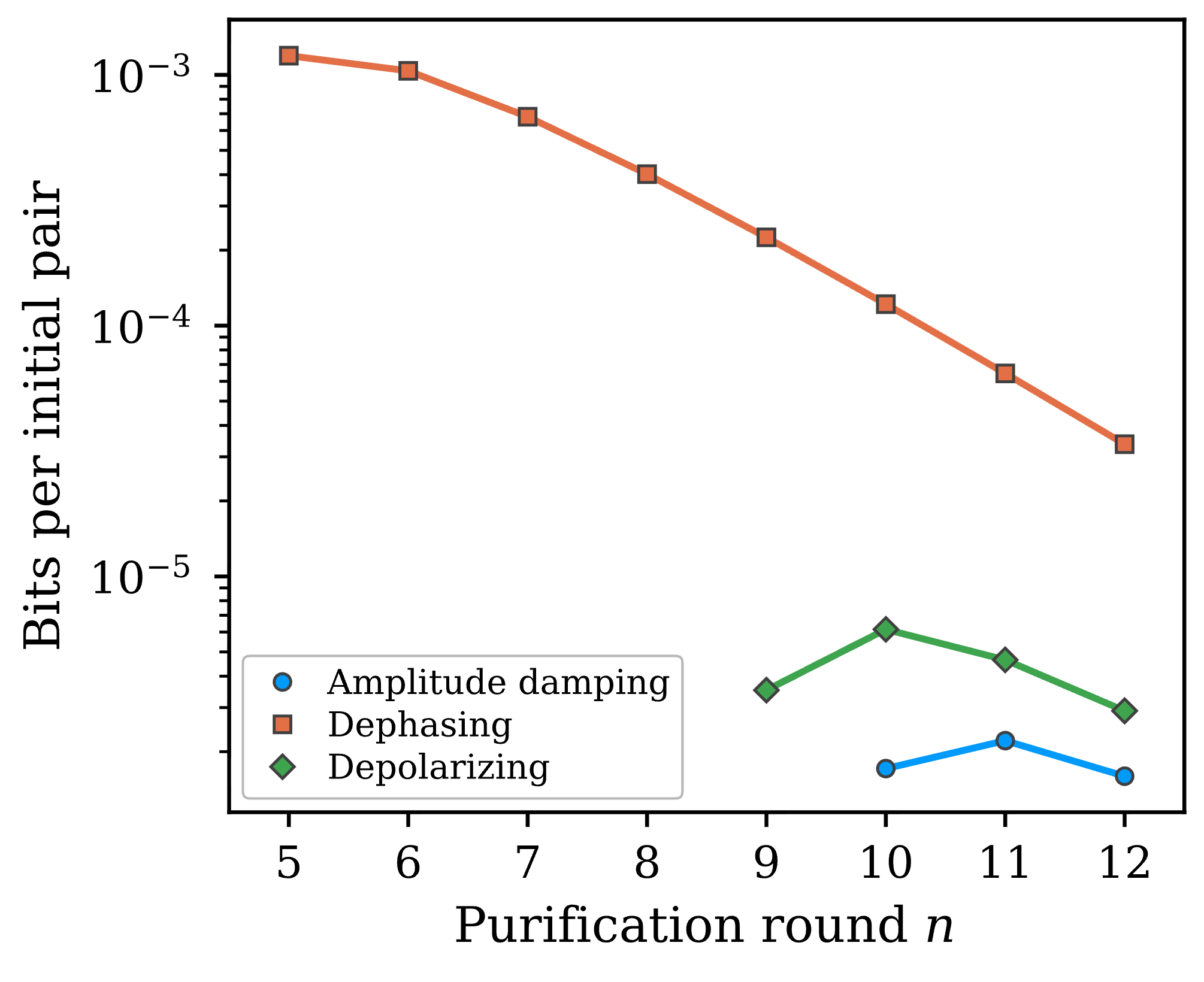}
        \caption{}
        \label{fig:keyrate_purificationb}
    \end{subfigure}
    \caption{(a) Secure key rate versus purification round at $L = 30$~km with $\eta_B = 0.9$. Purification enables positive key rates in regimes where unpurified states yield no secure key. (b) Effective rate
$r_{\rm eff}$ per initial pair, including the $2^n$ pair cost and the
cumulative success probability; only rounds with $r_{\rm eff}>0$ are shown on
the logarithmic scale.}
    \label{fig:keyrate_purification_combined}
\end{figure*}

\section{ENTANGLEMENT PURIFICATION FOR NOISE MITIGATION}
\label{purification}

Having established that realistic noise substantially impairs 1SDI-QKD performance, often eliminating secure key generation entirely, we investigate entanglement purification as a countermeasure. Purification does not diminish the efficiency advantage that motivates the
one-sided setting, since the detection threshold of a device-independent
protocol is a property of the apparatus rather than of the shared state: it
extends the tolerable channel noise but buys no additional loss budget. The
$50.1\%$ requirement of the two-setting protocol therefore remains well below
the $82.8\%$ of CHSH-based DI-QKD in the purified setting exactly as in the
ideal one, a margin of roughly $11$~km of standard fiber. Purification protocols distill high-fidelity entangled pairs from larger ensembles of noisy pairs, potentially restoring the correlations needed for secure key generation.

Section \ref{sec:framework} treats Bob's measurement apparatus as
an uncharacterised black box, whereas BBPSSW requires a CNOT gate and a
computational-basis measurement on each side. We therefore assume that
purification is performed by trusted entangling hardware co-located with the
source, upstream of the untrusted detection stage: the device-independence
assumption applies to Bob's \emph{detectors}, which are the components exposed
to the detection loophole and which set the $50.1\%$ efficiency threshold, not
to the entangling operations that prepare the state.

\subsection{The BBPSSW Purification Protocol}

We employ the BBPSSW protocol~\cite{bennett1996purification}, which operates on pairs of noisy entangled states to probabilistically produce a single higher-fidelity pair. The protocol proceeds as follows: (1) Alice and Bob each apply a CNOT gate between their respective qubits from two shared pairs, (2) both measure their target qubits in the computational basis, (3) they compare outcomes via classical communication, and (4) if outcomes agree, the control pair is kept with improved fidelity; otherwise both pairs are discarded.
For input states in Werner form, the output fidelity after one successful round is~\cite{bennett1996purification}:
\begin{equation}
    F' = \frac{F^2 + \frac{1}{9}(1-F)^2}{F^2 + \frac{2}{3}F(1-F) + \frac{5}{9}(1-F)^2}.
    \label{eq:bbpssw_fidelity}
\end{equation}
This recurrence has a stable fixed point at $F = 1$ and an unstable fixed point at $F = 1/2$. Consequently, for $F > 1/2$, repeated purification drives $F \to 1$, while for $F < 1/2$, purification fails.
The success probability for each round is:
\begin{equation}
    P_{\text{succ}}(F) = F^2 + \frac{2}{3}F(1-F) + \frac{5}{9}(1-F)^2.
    \label{eq:bbpssw_psucc}
\end{equation}
Because each round consumes two pairs to produce one and succeeds only with
probability $P_{\rm succ} < 1$, the throughput per \emph{initial} pair after $n$
rounds is $r_{\rm eff}(n) \;=\; \frac{r_\infty(F_n)}{2^{\,n}}
\\\prod_{k=0}^{n-1} P_{\rm succ}(F_k)$ where $F_k$ is the fidelity entering round $k$. We report both $r_\infty$, the
rate per successfully purified pair, and $r_{\rm eff}$, which carries the full
resource cost.
\subsection{Pauli Twirling for Non-Bell-Diagonal States}
The BBPSSW protocol assumes Bell-diagonal input states. However, amplitude damping produces non-Bell-diagonal states. Although twirling preserves the Bell-state fidelity, it can reduce other entanglement and correlation measures. Therefore, in our security analysis we compute the secret key rate from the measurement statistics of the twirled (Bell-diagonal) state used as input to purification. Before purification, such states must be converted to Bell-diagonal form via Pauli twirling:
\begin{equation}
    \mathcal{T}(\rho) = \frac{1}{4}\sum_{i \in \{I,X,Y,Z\}} (\sigma_i \otimes \sigma_i)\rho(\sigma_i \otimes \sigma_i).
    \label{eq:twirling}
\end{equation}
Twirling preserves fidelity but reduces concurrence for non-Bell-diagonal states. For amplitude damping at noise parameter $p = 0.3$, twirling reduces concurrence by approximately 18\% (from $C = 0.837$ to $C = 0.687$) while maintaining $F = 0.843$.

\begin{figure*}[t]
    \centering
    \begin{subfigure}[b]{0.45\textwidth}
        \centering
        \includegraphics[width=\textwidth]{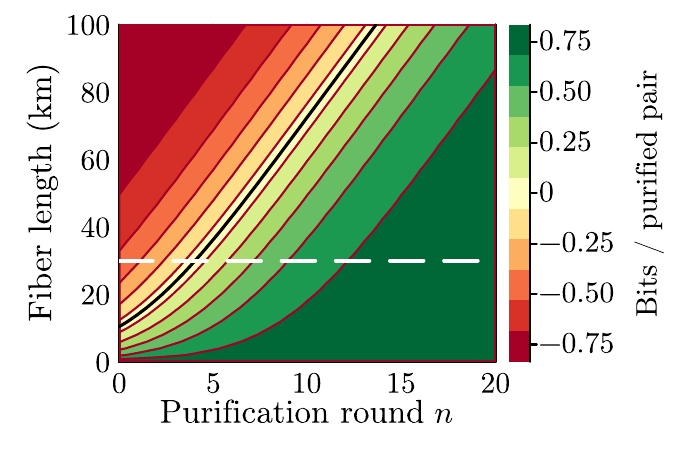}
        \caption{Dephasing ($L_c = 40$~km)}
    \end{subfigure}
    \hfill
    \begin{subfigure}[b]{0.45\textwidth}
        \centering
        \includegraphics[width=\textwidth]{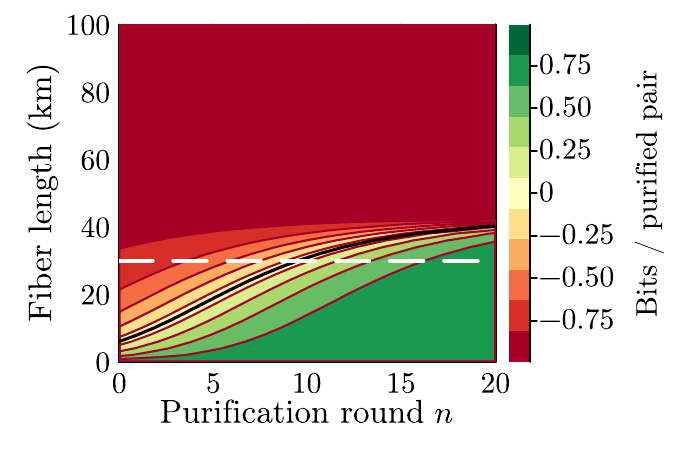}
        \caption{Amplitude damping ($L_c = 24$~km)}
    \end{subfigure}
    \hfill
    \begin{subfigure}[b]{0.45\textwidth}
        \centering
        \includegraphics[width=\textwidth]{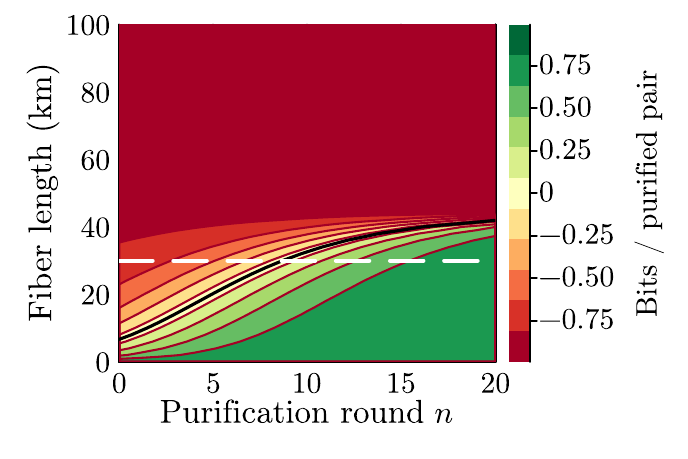}
        \caption{Depolarizing ($L_c = 40$~km)}
    \end{subfigure}
    \caption{Secure key rate versus fiber length and purification round. The zero-contour (black line) separates operational and non-operational regimes. The white dashed line marks $L = 30$~km.}
    \label{fig:contour_purification}
\end{figure*}

\subsection{Purification Results}
We present results for BBPSSW purification at fiber length $L = 30$~km. To model the distinct physical mechanisms underlying each noise type, we employ noise-type-specific coherence lengths: $L_c = 40$~km for dephasing, $L_c = 24$~km for amplitude damping, and $L_c = 40$~km for depolarizing noise~\cite{simon2010quantum}.


Figure~\ref{fig:fidelity_purification} shows the fidelity evolution over purification rounds. All noise types satisfy $F > 0.5$ at $L = 30$~km, ensuring purification convergence. Dephasing starts at $F \approx 0.74$ and reaches $F > 0.95$ after seven rounds.
Amplitude damping and depolarizing start at $F \approx 0.59$--$0.60$ and require
ten rounds to reach $F > 0.90$.

\begin{table}[t]
\centering
\caption{Implied noise strengths at $L=30$ km using the distance-dependent models.}
\begin{tabular}{lcc}
\hline
Noise type & $L_c$ (km) & Noise parameter at $L=30$ km \\
\hline
Dephasing & 40 & $p=\tfrac12(1-e^{-30/40})\approx 0.264$ \\
Amplitude damping & 24 & $\gamma=1-e^{-30/24}\approx 0.713$ \\
Depolarizing & 40 & $q=1-e^{-30/40}\approx 0.528$ \\
\hline
\end{tabular}
\end{table}



Figure~\ref{fig:keyrate_purification} shows the secure key rate versus
purification round. The critical observation is that purification converts
\emph{impossible} key generation ($r \leq 0$) into \emph{possible} ($r > 0$).
All three channels begin with negative key rates at $L = 30$~km. Dephasing,
starting from the highest initial fidelity, crosses into the secure regime
after five rounds, whereas amplitude damping and depolarizing require
approximately nine rounds owing to their lower initial fidelities
($F \approx 0.58$--$0.60$, Fig.~\ref{fig:fidelity_purification}).



Figure~\ref{fig:contour_purification} presents contour plots of secure key rate as a function of fiber length and purification round. The zero-contour (black line) separates regions where positive key generation is possible from those where it is not. Dephasing permits the largest secure region, while depolarizing restricts operation to shorter distances. Beyond approximately ten rounds, expansion of the secure region saturates.

Table~\ref{tab:purification_strategy} summarizes operational guidelines. The BBPSSW protocol consumes $2^n$ initial pairs per output pair after $n$ rounds.


\begin{table}[htbp]
\centering
\caption{Maximum fiber length (km) supporting a positive asymptotic key rate at
$\eta_B=0.9$, as a function of purification depth $n$.}
\label{tab:purification_strategy}
\begin{tabular}{lcccc}
\hline
\textbf{Channel} & $n=0$ & $n=2$ & $n=5$ & $n=10$ \\
\hline
Dephasing         & 10.4 & 18.6 & 36.3 & 72.6 \\
Amplitude damping &  6.0 & 10.4 & 18.8 & 30.9 \\
Depolarizing      &  6.6 & 11.4 & 20.3 & 32.7 \\
\hline
\end{tabular}
\end{table}


\begin{algorithm*}[t]
\small
\DontPrintSemicolon
\SetAlgoLined
\caption{Numerical pipeline for the asymptotic key rate and its purified variant}
\label{alg:pipeline}
\KwIn{$\theta=\pi/4$; channel $\mathcal{E}_\lambda\in\{\mathcal{E}_{\mathrm{deph}},
      \mathcal{E}_{\mathrm{ad}},\mathcal{E}_{\mathrm{depol}}\}$ of strength $\lambda$;
      $\eta_B$; Gauss--Radau order $m=15$; NPA level $\ell=1$;
      purification depth $N$ (Section~5 only)}
\KwOut{$r_\infty$; and, for the purified variant,
       $\{F_n\}_{n=0}^{N}$, $r_\infty(F_n)$ and $r_{\mathrm{eff}}(n)$}
\BlankLine
\tcp{State preparation and channel}
$\rho_{AB}\gets\ket{\psi(\theta)}\!\bra{\psi(\theta)}$, \quad
$\ket{\psi(\theta)}=\cos\theta\ket{00}+\sin\theta\ket{11}$\;
$\rho'_{AB}\gets\sum_k (K_k\otimes\mathbb{1})\,\rho_{AB}\,(K_k\otimes\mathbb{1})^{\dagger}$
  \tcp*{$(\mathcal{E}_\lambda\otimes\mathbb{1})$, Eq.~(24); Alice is the first tensor factor}
\tcp{Measurements and observed statistics}
Alice: $M^{(1)}_{a}\gets\{\ket{0}\!\bra{0},\ket{1}\!\bra{1}\}$,\;
$M^{(2)}_{a}\gets\{\ket{+}\!\bra{+},\ket{-}\!\bra{-}\}$
  \tcp*{Eqs.~(7)--(8); trusted, fair sampling}
Bob: $M^{(y,\eta_B)}_{b}\gets\eta_B\,\Pi_{b|y}$ for $b\in\{1,2\}$,\;
$M^{(y,\eta_B)}_{\emptyset}\gets(1-\eta_B)\mathbb{1}$
  \tcp*{Eq.~(9)}
$p(a,b|x,y)\gets\Tr\!\big[(M^{(x)}_{a}\otimes M^{(y,\eta_B)}_{b})\,\rho'_{AB}\big]$,
  $a,x,y\in\{1,2\}$, $b\in\{1,2,\emptyset\}$
  \tcp*{Eq.~(25)}
$H(A_1|B_1)\gets-\sum_{a,b}p(a,b|1,1)\,\log_2\!\big[p(a,b|1,1)/p(b|1)\big]$
  \tcp*{Eq.~(3)}
\tcp{$H(A_1|E)$ via Brown--Fawzi--Fawzi on the NPA hierarchy}
$\{t_i,w_i\}_{i=1}^{m}\gets$ Gauss--Radau nodes; discard $i=1$\;
$\alpha_i\gets\tfrac{3}{2}\max\!\big(t_i^{-1},(1-t_i)^{-1}\big)$\;
Build moment matrix $\Gamma$ at level $\ell$ over $\{\Pi^A,\Pi^B,Z,Z^{\ast}\}$,
  extended by the localizing set of $A_1A_2+A_2A_1=0$\;
Impose $\langle\Pi^A_{a|x}\Pi^B_{b|y}\rangle=p(a,b|x,y)$ for the 12 retained
  $(a,b,x,y)$, and $\langle Z^{\ast}Z\rangle,\langle ZZ^{\ast}\rangle\le\alpha_i$\;
\For{$i\gets 1$ \KwTo $m-1$}{
  $o_i\gets\min\limits_{\Gamma\succeq 0}\;\dfrac{w_i}{t_i\ln 2}\Big[$\;
  \Indp
  $\sum_{a}\big\langle Z_a+Z_a^{\ast}+(1-t_i)Z_a^{\ast}Z_a+t_i Z_aZ_a^{\ast}\big\rangle\Big]$
    \tcp*{MOSEK, rel.\ gap $10^{-6}$, feas.\ $10^{-7}$}
  \Indm
}
$H(A_1|E)\gets\sum_{i}\dfrac{w_i}{t_i\ln 2}+\sum_i o_i$\;
$r_\infty\gets H(A_1|E)-H(A_1|B_1)$
  \tcp*{Eq.~(26)}
\tcp{BBPSSW at fibre length $L$ (Section~5)}
$\lambda\gets\lambda(L)$ by Eqs.~(17),(20),(23); recompute $\rho'_{AB}$\;
\lIf{$\rho'_{AB}$ is not Bell-diagonal}{$\rho'_{AB}\gets\mathcal{T}(\rho'_{AB})$ \tcp*{Eq.~(40)}}
$F_0\gets\bra{\Phi^{+}}\rho'_{AB}\ket{\Phi^{+}}$\;
$\rho_{F_0}\gets F_0\ket{\Phi^{+}}\!\bra{\Phi^{+}}
   +\dfrac{1-F_0}{3}\!\!\sum_{\Psi\neq\Phi^{+}}\!\!\ket{\Psi}\!\bra{\Psi}$
  \tcp*{Werner projection; see Section~5.2}
\For{$n\gets 1$ \KwTo $N$}{
  $F_n\gets\big[F_{n-1}^{2}+\tfrac{1}{9}(1-F_{n-1})^{2}\big]\big/P_{\mathrm{succ}}(F_{n-1})$
    \tcp*{Eq.~(38)}
  $r_{\mathrm{eff}}(n)\gets 2^{-n}\,r_\infty(F_n)
    \prod\limits_{k=0}^{n-1}P_{\mathrm{succ}}(F_k)$,\quad
  $P_{\mathrm{succ}}(F)= F^{2}+\tfrac{2}{3}F(1-F)+\tfrac{5}{9}(1-F)^{2}$
    \tcp*{Eq.~(39); $r_\infty(F_n)$ from lines 5--12 with $\rho'_{AB}=\rho_{F_n}$}
}
\end{algorithm*}

\section{Conclusion and Future Work} \label{conc}
Situated between the stringent requirements of fully DI-QKD (which demands $>$82\% efficiency and loophole-free Bell tests) and the partial trust assumptions of MDI-QKD (which requires trusted state preparation), one-sided device-independent QKD (1SDI-QKD) offers a pragmatic intermediate: true device independence on the untrusted side with substantially relaxed efficiency requirements. In this paper, we present a comprehensive noise analysis of 1SDI-QKD, extending the 1SDI-QKD framework of Masini and Sarkar~\cite{masini2026one} to realistic quantum channels. Our work addresses a critical gap between theoretical security proofs, which assume idealized conditions, and the practical deployment of 1SDI-QKD over noisy fiber-optic networks.

Our analysis establishes a clear noise hierarchy: dephasing tolerates the
greatest channel noise before security is lost ($p_{\rm crit}\approx0.50$,
against $0.449$ for amplitude damping and $0.220$ for depolarizing), while
depolarizing noise imposes the most stringent detection-efficiency requirements,
exceeding $97\%$ at $20\%$ noise where amplitude damping requires $76\%$ and
dephasing $85\%$. 
Crucially, the transition to insecurity is not synonymous with the disappearance of quantum correlations: we identify parameter regimes in which the shared state remains \emph{steerable}, yet the asymptotic key rate is already non-positive, $r_\infty \le 0$. This occurs because noise can increase Bob's conditional entropy $H(A_1|B_1)$ sufficiently to overwhelm the bound on Eve's uncertainty, so that steering alone does not certify operational key generation in the \emph{unpurified} 1SDI-QKD setting. Consistent with this, security can vanish while substantial entanglement persists (e.g., $C \approx 0.68$--$0.80$ at the security threshold), demonstrating that nonzero entanglement, and even residual steerability, is not by itself a sufficient design target. Instead, positive key rates require \emph{strong enough} steering correlations. When entanglement purification is available, this picture changes: purification increases the effective Bell-state fidelity and can restore sufficiently strong steering, making the two-setting benchmark $S_2 > 1/\sqrt{2}$ operationally meaningful as a threshold for recovering $r_\infty>0$ in regimes that are otherwise insecure without purification.

The BBPSSW entanglement purification protocol effectively restores security in otherwise insecure regimes, but our resource analysis shows that the benefit is
strongly channel-dependent. Accounting for both the exponential pair consumption
($2^n$) and the cumulative success probability, the effective rate (Fig. \ref{fig:keyrate_purificationb}) peaks at five
rounds for dephasing ($\sim\!10^{-3}$ bits per initial pair) and at ten or more
for the isotropic channels, where it reaches only $\sim\!10^{-6}$. Purification
therefore acts principally as a range-extension mechanism rather than a
rate-recovery one: five rounds extend dephasing-limited operation from
$10.4$~km to $36.3$~km at $\eta_B = 0.9$, whereas amplitude damping and
depolarizing reach only $18.8$~km and $20.3$~km at the same depth and saturate
near $42$~km, where the post-channel fidelity falls below the BBPSSW convergence
threshold $F = 0.5$. Metropolitan-scale operation is thus attainable where phase
noise dominates; for isotropic-noise-limited links, quantum repeaters rather
than deeper purification are the appropriate next step.

These findings establish quantitative boundaries for practical 1SDI-QKD deployment and demonstrate that noise-aware entanglement management provides a viable path from theoretical security to real-world quantum key distribution.


Several directions warrant future investigation: (i)~finite-key security analysis for practical block sizes, where statistical fluctuations may further constrain operating regimes \cite{tomamichel2012tight}; (ii)~experimental validation using state-of-the-art photon sources and superconducting nanowire detectors \cite{reddy2020superconducting}; (iii)~alternative purification protocols such as DEJMPS \cite{deutsch1996quantum} or hashing-based schemes that may offer improved resource efficiency,  (iv)~integration with quantum repeater architectures for range extension beyond metropolitan scales \cite{dur1999quantum}; and (v) device-independent certification of the purification step itself, which
would remove the trusted-hardware assumption made in Section \ref{purification}.

\section*{CRediT authorship contribution statement}

\textbf{Syed M. Arslan:} Conceptualization, Methodology, Writing -- original draft. \textbf{Muhammad T. Rahim:} Methodology, code,  Validation, Writing -- review \& editing. \textbf{Asad Ali:} Formal analysis, Writing -- review \& editing. \textbf{Hashir Kuniyil:} Investigation, Visualization. \textbf{Saif Al-Kuwari:} Supervision, Project administration.







\bibliographystyle{elsarticle-num}
\bibliography{ref}

\end{document}